\documentclass[12pt,preprint]{aastex}
\usepackage{color}

\slugcomment{Revised on 2011 September 20}

\shorttitle{Interpretation of (596) Scheila's Triple Dust Tails}
\shortauthors{Ishiguro et al.}

\begin{document}

\title{Interpretation of (596) Scheila's Triple Dust Tails}


\author{Masateru \textsc{Ishiguro}\altaffilmark{1}}
\affil{Department of Physics and Astronomy, Seoul National University,
Gwanak, Seoul 151-742, South Korea}

\author{Hidekazu \textsc{Hanayama}}
\affil{Ishigakijima Astronomical Observatory, National Astronomical
Observatory of Japan, Ishigaki, Okinawa, 907-0024, Japan}

\author{Sunao \textsc{Hasegawa} and Yuki \textsc{Sarugaku}}
\affil{Institute of Space and Astronautical Science (ISAS),
JAXA, Sagamihara, Kanagawa 252-5210,
Japan}

\author{Jun-ichi \textsc{Watanabe}\altaffilmark{2}}
\affil{National Astronomical Observatory of Japan, Mitaka, Tokyo,
181-8588, Japan}

\author{Hideaki \textsc{Fujiwara} and Hiroshi \textsc{Terada}}
\affil{Subaru Telescope, National Astronomical Observatory of Japan,
Hilo, HI 96720, USA}

\author{Henry H. \textsc{Hsieh}}
\affil{Institute for Astronomy, University of Hawaii, Honolulu, HI 96822, USA}

\author{Jeremie J. \textsc{Vaubaillon}}
\affil{Observatoire de Paris, I.M.C.C.E., Denfert Rochereau, Bat. A.,
FR-75014 Paris, France} 

\author{Nobuyuki \textsc{Kawai}}
\affil{Department of Physics, Tokyo Institute of Technology, Meguro,
Tokyo 152-8551, Japan} 

\author{Kenshi \textsc{Yanagisawa} and Daisuke \textsc{Kuroda}}
\affil{Okayama Astrophysical Observatory, National Astronomical
Observatory of Japan, Asaguchi, Okayama 719-0232, Japan}

\author{Takeshi \textsc{Miyaji}\altaffilmark{2} and Hideo
\textsc{Fukushima}\altaffilmark{2}}
\affil{National Astronomical Observatory of Japan, Mitaka, Tokyo,
181-8588, Japan} 

\author{Kouji \textsc{Ohta}}
\affil{Department of Astronomy, Kyoto University, Sakyo, Kyoto 606-8502, Japan}

\author{Hiromi \textsc{Hamanowa}}
\affil{Hamanowa Astronomical Observatory, Motomiya, Fukushima 969-1204, Japan}

\author{Junhan \textsc{Kim}}
\affil{Yangcheon-gu, Seoul, South Korea}

\author{Jeonghyun \textsc{Pyo}}
\affil{Korea Astronomy and Space Science Institute (KASI), Daejeon
305-348, South Korea}

\and

\author{Akiko M. \textsc{Nakamura}}
\affil{Department of Earth and Planetary Sciences, Kobe University,
Nada, Kobe 657-8501, Japan}


\altaffiltext{1}{To whom correspondence should be addressed. E-mail:
ishiguro@snu.ac.kr}
\altaffiltext{2}{Ishigakijima Astronomical Observatory, National Astronomical
Observatory of Japan, Ishigaki, Okinawa, 907-0024, Japan}

\begin{abstract}
Strange-looking dust cloud around asteroid (596) Scheila was discovered
on 2010 December 11.44--11.47. Unlike normal cometary tails, it consisted of
three tails and faded within two months. We constructed a model to
reproduce the morphology of the dust cloud based on the laboratory
measurement of high velocity impacts and the dust dynamics. As the
result, we succeeded in the 
reproduction of  peculiar dust cloud by an impact-driven ejecta plume
consisting of an impact cone  and downrange plume.
Assuming an impact angle of 45\arcdeg~, our model suggests
that a decameter-sized asteroid collided with (596) Scheila from the
direction of ($\alpha_{im}$, $\delta_{im}$) = (60\arcdeg~, -40\arcdeg~)
in J2000 coordinates
on 2010 December 3. The maximum ejection velocity of the dust particles
exceeded 100 m/s.  Our results suggest that
the surface of (596) Scheila consists of materials with low tensile
strength.
\end{abstract}

\keywords{comets: general -- comets: individual ((596) Scheila)
minor planets, asteroids --- general}

\section{Introduction}

In this paper, we suggest one plausible explanation for the peculiar
dust cloud of (596) Scheila.

(596) Scheila is a large asteroid (113--120 km in diameter) orbiting the
Sun in the outer region of the main belt with the orbital period of 5.01
years \citep{Tedesco2002,Usui2011}. The orbital elements of the object
are typical for the outer main-belt, that is, the semi-major axis, the
eccentricity and the inclination are 2.93AU, 0.164, and 14.7\arcdeg. 
An unexpected dust cloud of (596) Scheila was discovered on
2010 December 11.4 with the 0.68-m f/1.8 Schmidt telescope
\citep{Larson2010}. The observation with the same instruments on 2010
December 3.4 showed a diffuseness at magnitude 13.2, about 1.3
mag brighter than that of the observation in the previous month
\citep{Larson2010}. \citet{Ishiguro2011} found that the dust particles
ranging from 0.1--1 \micron~ to 100 \micron~ were ejected impulsively on
2010 December 3.5 $\pm$1.0 through the synchrone analysis of extended
dust structure appeared after 2011 February. It is therefore likely that 
\citet{Larson2010} observed (596) Scheila immediately after the dust
emission.  The total mass of the ejecta was estimated to be (0.2--6)
$\times$ 10$^8$ kg, depending on the assumed particle size and the mass
density \citep{Jewitt2011,Hsieh2011,Bodewits2011,Ishiguro2011}. To date,
gas emission has never been detected \citep{Jewitt2011,Hsieh2011,Bodewits2011}. 
Accordingly, it is natural to think that the comet-like activity was
triggered by an impact. The impactor diameter was estimated to be 20--50 m
\citep{Jewitt2011,Bodewits2011,Ishiguro2011}. 

Similar evidence for asteroid--asteroid impact was reported in another
object. The dust cloud of P/2010 A2 (LINEAR) was discovered on 2010
January 6, showing not only a comet-like extended dust cloud but also a
mysterious X-shaped debris pattern \citep{Jewitt2010}. It has been
proposed that the dust ejecta was created by the impact of a small
objects in 2009 February or March (about 10 months before the
discovery), although it cannot be ruled out that the asteroid's
rotational spin-up resulted in a mass loss that formed a comet-like
debris tail \citep{Snodgrass2010,Jewitt2010}. 

The morphology of (596) Scheila's dust cloud was also mysterious, in
that it consisted of three prominent structures. Figure ~\ref{fig1} (a)
shows the image taken on 2010 December 12 at Ishigakijima Astronomical
Observatory with 1-m telescope and a 3ch (the g', Rc, and Ic-bands)
simultaneously imaging system. In the image, three
components appear: the northern tail, southern tail, and westward tail
\citep{Ishiguro2011}. Similar structures were found in \citet{Jewitt2011}.
To date, no studies have addressed the physical implication of the
mysterious morphologies of impact-triggered dust clouds.
In this paper, we attempted to reconstruct the observed morphology of
(596) Scheila's triple dust tails on the basis of the impact hypothesis.
The image taken on 2010 December 12 at Ishigakijima
Astronomical Observatory was compared to the model which considers
the laboratory measurement of high velocity impact and dust dynamics,
and derive the best-fit parameters.

\section{Model Description}

\begin{figure}
\epsscale{0.80}
\plotone{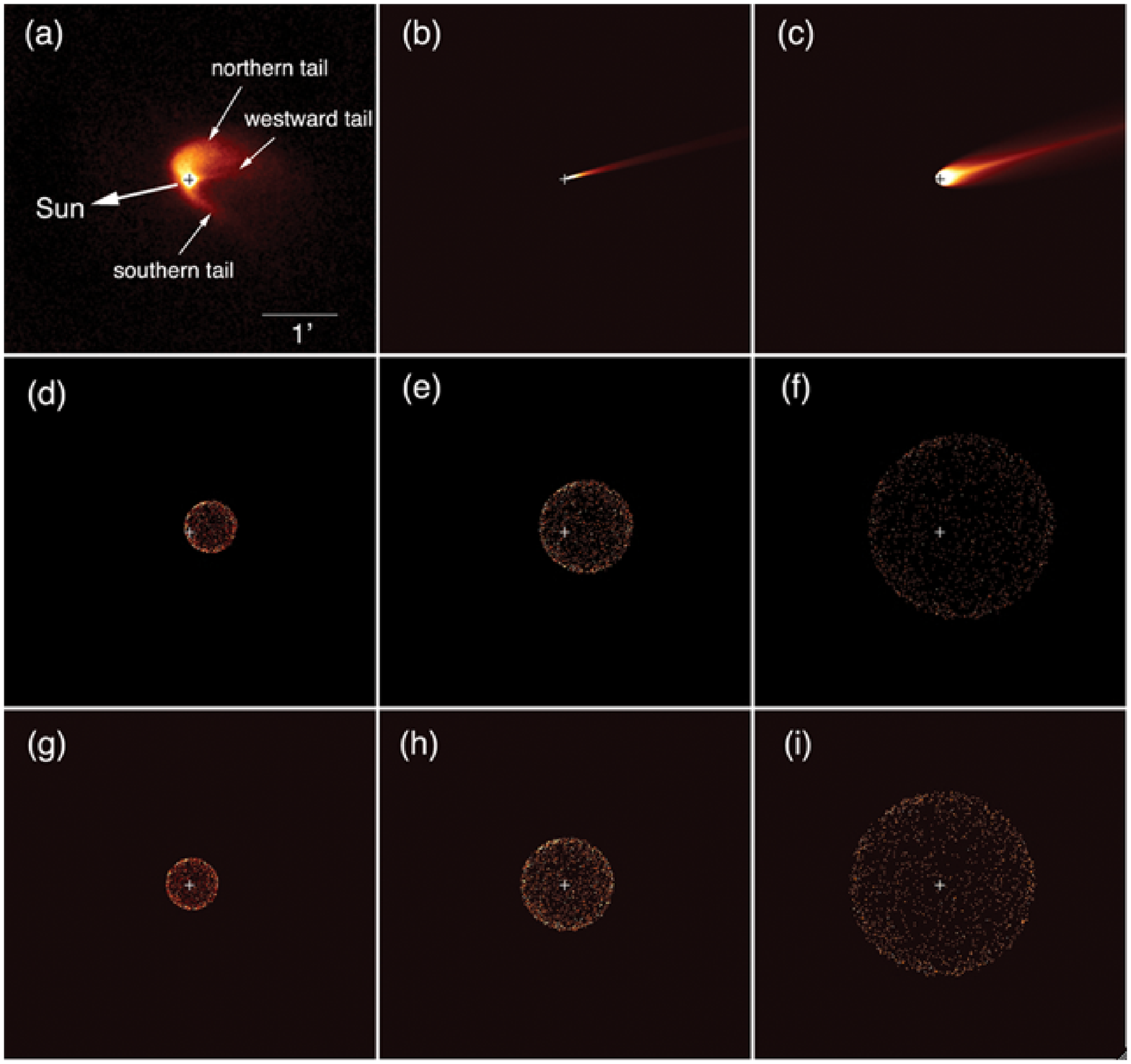}
\caption{(a) Image observed on 2010 December 12 at Ishigakijima Astronomical
Observatory. (b)--(i) Images of cometary dust ejection models for the
observation on 2010 December 12. (a)(b) Results of the continuous dust
ejection models using  parameters from 238P/Read \citep{Hsieh2009} and
22P/Kopff \citep{Ishiguro2007}. (d)--(i) The  results of an impulsive
isotropic dust ejection. We assumed 1 \micron-particles with the terminal
velocities of (d) 57 m/s, (e) 100 m/s, and (f) 200 m/s, and 10
\micron-particles with the terminal velocities of (g) 57 m/s, (h) 100 
m/s, and (i) 200 m/s.  In all panels, the  emission  source, (596)
Scheila, is at the center of each image, and the field of view is
3.5\arcmin $\times$ 5.8\arcmin. 
 \label{fig1}}
\end{figure}

As a beginning we would like to clarify difference in shape between
normal comets and (596) Scheila. The morphology of (596) Scheila cannot
be explained by the sublimation of ice. Figure ~\ref{fig1} (b) and (c)
show the results of model simulations on 2010 December 12 performed
under the assumption of comet-like dust ejection. We applied continuous 
dust-ejection models using the parameters of 238P/Read as an analog of a
main-belt comet that was activated by the sublimation of ice
\citep{Hsieh2009} (Figure ~\ref{fig1} (b)), and those of 22P/Kopff as an
analog of a Jupiter-family comet \citep{Ishiguro2007} (Figure
~\ref{fig1} (c)).
In these models, it is assumed that dust particles are ejected in
cone-shape jets that are radially symmetric with respect to the
Sun-object axis with a half-opening angle of 45\arcdeg~ (238P/Read) and
60\arcdeg~ (22P/Kopff).  We considered continuous dust emission from two
months prior to the observation on 2010 December 12.
In these model images, the dust cloud smoothly extended in an almost
anti-solar direction. There is only one dust tail in these images.
The observed image differs from these simulation images in that it
consisted of multiple tails. 

Secondly, we show simple impulsive emission models in Figure ~\ref{fig1}
(d)--(i). In these models, we assumed the isotropic dust emission of 1
\micron~ and 10\micron-particles with different velocities ejected on
2010 December 3. We adopted the ejection day based on the previous
studies; the dust emission should have occurred on 2010 December
3.5$\pm$1.0 \citep{Ishiguro2011} but before 2010 December 3.4
\citep{Larson2010}. It may seem at a first glance that there should be
dust particles ejected with high terminal velocity ($v_{tml}\approx100$
m/s or higher) because the rim diameters of models with the terminal
velocity of $<$ 100 m/s look smaller than that of the observed image. In
addition, small grains ($\lesssim$1\micron) should exist 
because the dust cloud was deflected toward the anti-solar direction by
solar radiation pressure.
For comparison, assuming the mass density of 1670 kg/m$^3$ (equivalent
to the mass density of the Tagish Lake meteorite
\citet{Hiroi2003,Zolensky2002}), the escape velocity from (596) Scheila
is 55 m/s. The estimated terminal velocity is two times faster than
the escape velocity.

\begin{figure}
\epsscale{0.80}
\plotone{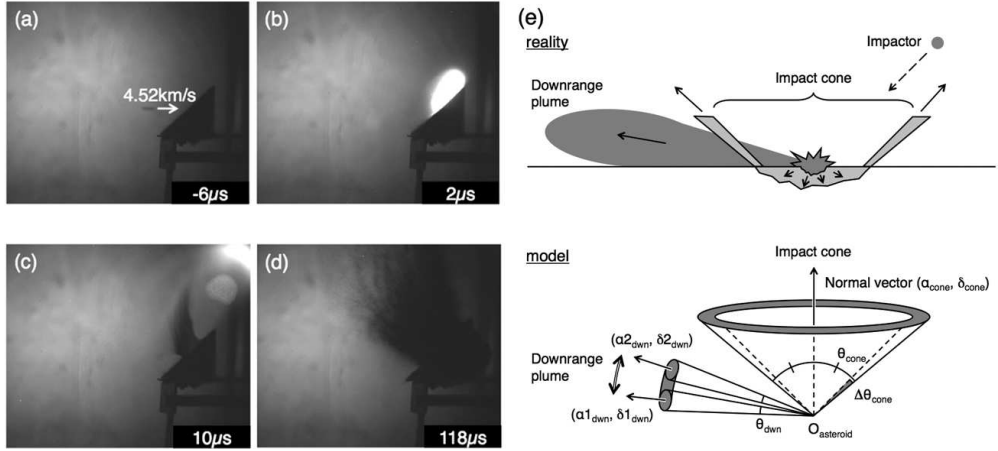}
\caption{Sample images of the laboratory impact experiment. The
 experiment was conducted with a two-stage light gas gun at the
 Institute of Space and Astronautical Science (ISAS), Japan Aerospace
 Exploration Agency (JAXA). A 3.2-mm steel sphere was accelerated to
 4.521 km/s with the gun using a sabot \citep{Kawai2010}. The material of
 the target was serpentine. The 
 chamber of the target was evacuated to a pressure of 7 Pa. The time
 resolution of the high-speed camera was 4 $\mu$s. The number at the bottom
 right of each frame shows the elapsed time, t, in microseconds, with
 $t$=0 at the impact. (a) An image before the impact. (b) Impact of the
 projectile forming a luminous downrange plume. (c) Dust particles were
 ejected from the target while the luminous downrange plume passed with
 high speed. (d) The growth of the conical ejecta curtain.
(e) Schematic diagram showing how our simulation modeled the impact
 phenomena. Impact ejecta consist of a conical ejecta curtain and
 downrange plume.
 \label{fig2}}
\end{figure}

To reconstruct the observed morphology of the (596) Scheila dust cloud,
we considered a new dust emission model, described below. We
assumed that the dust particles were ejected in two different forms from
an impact point, i.e., a conical impact ejecta curtain and a downrange 
plume, which are commonly observed in oblique impact experiments.
Figure ~\ref{fig2} shows an example of the laboratory oblique impact
experiment. It was conducted with a two-stage light gas gun at
the Institute of Space and Astronautical Science (ISAS), Japan Aerospace
Exploration Agency (JAXA). A 3.2-mm steel sphere was accelerated with
the gun in a sabot \citep{Kawai2010} to 4.521 km/s, which is typical
impact velocity in outer main-belt \citep{O'Brien2011}. Soon after the
impact, the dust particles were ejected as the luminous downrange plume
with high speed in the horizontal direction with respect to the local
surface. Later, the conical ejecta 
curtain was grown. We simplified these forms, as shown in Figure
~\ref{fig2} (e). The conical impact ejecta curtain is
symmetrical with respect to a vector normal to the asteroid surface
($\alpha_{cone}$, $\delta_{cone}$) with a half-opening angle of
$\theta$. Dust particles are assumed to be ejected between
$\theta$-$\Delta \theta/2$ and $\theta$+$\Delta \theta/2$ (see the
shadowed area in Figure ~\ref{fig2} (e)). We modeled the downrange plume
as a stretched cone with a central axis ranging from ($\alpha1_{dwn}$,
$\delta1_{dwn}$) to ($\alpha2_{dwn}$,  $\delta2_{dwn}$). 

The dust particles ejected with the ejection velocity of $v_{ej}$ would
decelerate by the asteroid's gravity and reach the terminal velocity
$v_{tml}$. We applied the power law function of the ejection velocity of
dust particles:  

\begin{equation}
\left \{
\begin{array}{lclc}
v_{ej} & = & V_0 a^k & \\
v_{tml} & = & \sqrt{v_{ej}^2 - \frac{2GM_{596}}{R_{596}}} & ~~\left(v_{ej}>\sqrt{\frac{2GM_{596}}{R_{596}}}\right)
\end{array}
\right .
\end{equation}

\noindent
where $V_0$ is the reference ejection velocity (m/s) of the particles radius
$a$=1$\times$10$^{-6}$ (m), $k$ is the power index of size dependence of
the ejection velocity, and $M_{596}$ and $R_{596}$ are the mass and radius
of (596) Scheila. We considered the energy conservation for the terminal
velocity $v_{tml}$ in the second equation of Eq. (1).

An power-law size distribution with index $q$ was 
used. The number of dust particles within a size range of $a$ and
$a$+$da$ is given by: 

\begin{equation}
N(a)da =
\left \{
\begin{array}{lcl}
N_0 a^q~ da & \hspace{5mm} \mathrm{for} & a_{min}\leq a \leq a_{max} \\
0              & \hspace{5mm} \mathrm{for} & a < a_{min}, a > a_{max}
\end{array}
\right .
\end{equation}

\noindent
where $a_{max}$ and $a_{min}$ are the maximum and minimum particle sizes,
respectively, and $N_0$ represents the reference dust production rate.
We fixed $a_{min}$=0.1 \micron~ and $q$=-3.5 (discuss later),
because particles much smaller than the
wavelength are inefficient scatterers in the optical wavelength.
The size distribution exponent was also fixed to $q$=-3.5
\citep{Dohnanyi1969}. The maximum size $a_{max}$ can be derived when the
size-dependent velocity $v_{ej}$ becomes equal to the escape velocity
from (596) Scheila. 

The trajectories of the particles were computed from the terminal
velocity and the ratio of the force exerted by the solar radiation
pressure and the solar gravity ($\beta$).
It can be expressed as $\beta$=$K Q_{pr}/ \rho a$ , where
$K$ = 5.7$\times$10$^{-4}$ kg m$^{-2}$, and $Q_{pr}$ is the
radiation pressure coefficient averaged over the solar spectrum
\citep{Burns1979}. We assumed $Q_{pr}$=1. 
$\rho$ denotes the mass density of dust particles. We supposed that the
mass densities of the dust particles was 1670 kg/m$^3$. 
The model images were reconstructed using the Monte Carlo
approach for the parameters above
\citep{Ishiguro2007,Ishiguro2008,Sarugaku2007,Hsieh2009}.
We calculated the positions of dust particles at a given time by solving
the Keplerian equation. We considered the impulsive dust emission on
2010 December 3 as stated above.
The free parameters of our model are listed in Table 1.

\section{Results and Discussion}

\begin{figure}
\epsscale{1.0}
\begin{center}
\plotone{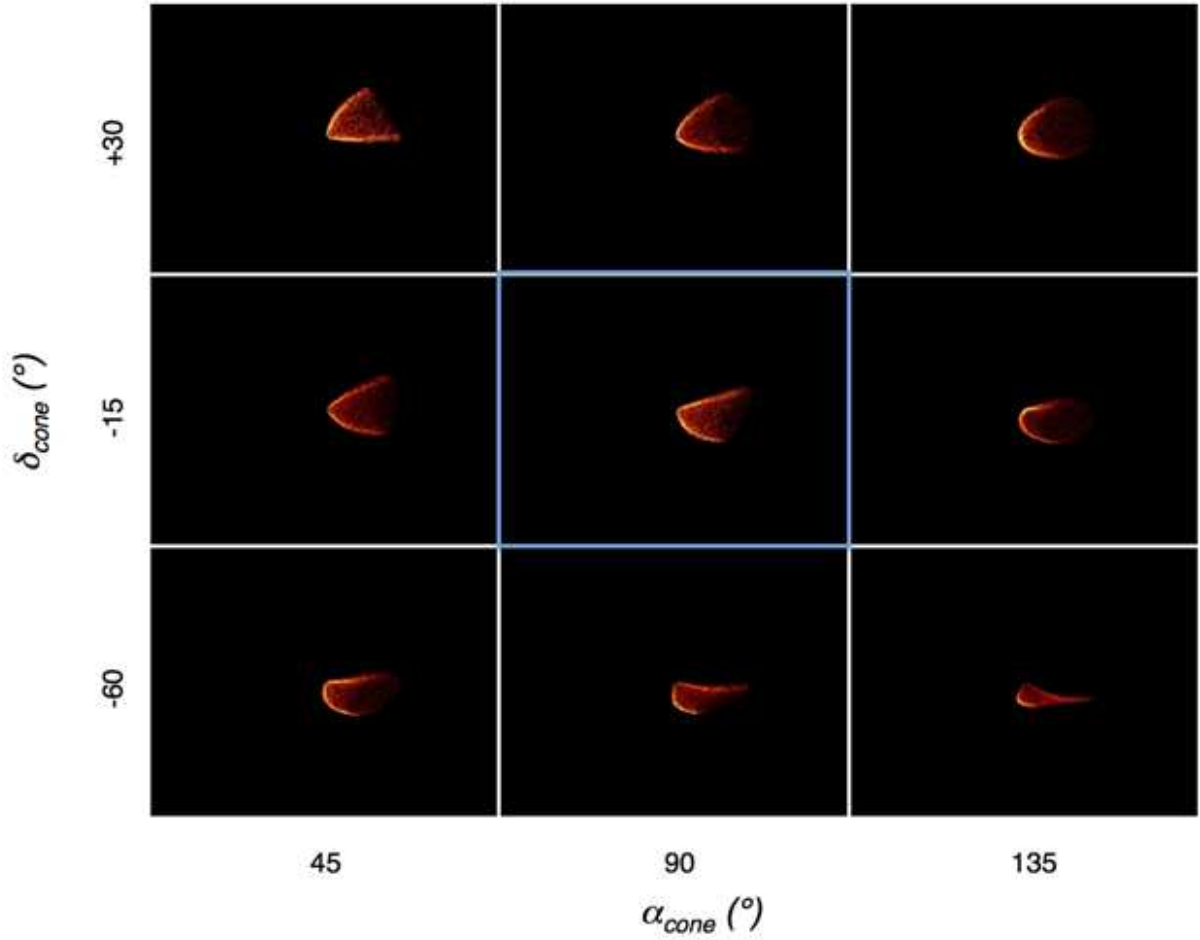}
\end{center}
\caption{Images of conical curtain ejection models for December 12, 2010
 for $\theta$=50\arcdeg~, $\Delta \theta$=10\arcdeg~, $V_0$=80 m/s, and
 different jet directions as labeled,  where $\alpha_{cone}$ is constant
 for each row of models and  $\delta_{cone}$ is constant   for each
 column of models. 
 In all panels,  the emission source, (596)
 Scheila, is at the  center of each  image.\label{fig3}}
\end{figure}

\begin{figure}
\epsscale{0.90}
\plotone{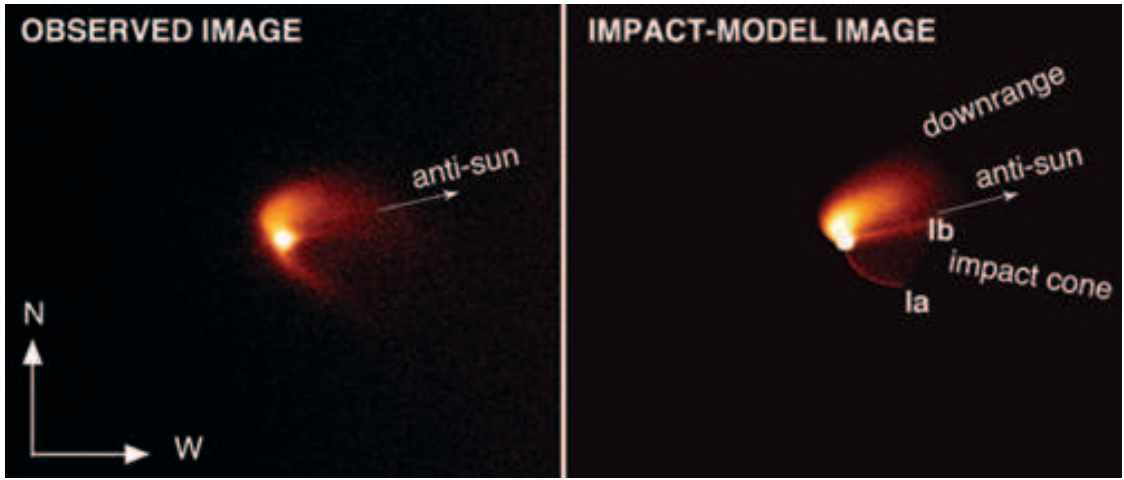}
\caption{Result of numerical simulation reproducing the observed
 morphology on 2010 December 12. In the model, we considered dust
 ejecta consisting of an impact cone and downrange plume. Filled circle
 indicates the asteroid's position; arrow indicates the anti-solar
 direction. The northern and southern tails were swept back by
 radiation pressure from the Sun, while the westward tail extended
 linearly because the orientation was close the anti-solar direction.
 \label{fig4}}
\end{figure}

Multiple simulations are carried out using various parameter sets, and
the resulting model images are then visually compared to the data to
find plausible model parameters. As the result, we obtained the best-fit values.
Figure ~\ref{fig3} shows some
example results for the conical ejecta curtain. First, we noticed that
the southern tail and westward tail (see 
figure ~\ref{fig1}) could be reproduced by a conical ejecta curtain
given a half-opening angle of ~50\arcdeg~. Our observations were
consistent with the case in which the central axis at ($\alpha_{cone}$,
$\delta_{cone}$) of the downrange was (90\arcdeg, -15\arcdeg) in the
J2000 coordinate system. 
Similarly, we examined the dependence of the downrange plume
($\alpha_{dwn}$,  $\delta_{dwn}$) on the central axis.
We notice the integral along the great circle joining from 
($\alpha1_{dwn}$,  $\delta1_{dwn}$) = (150\arcdeg, +40\arcdeg) to 
($\alpha2_{dwn}$,  $\delta2_{dwn}$) = (180\arcdeg, +50\arcdeg) matches
the observed image. 
As $\theta$  increased, the opening angle of   
the ejecta became broader; as $V_0$ increased, the dust cloud extended more
widely. Accordingly, these two variables, $\theta$ and $V_0$, were well
determined by comparison to the observed images. We estimated
$\theta$=15\arcdeg~ (the downrange plume), $\theta$=50\arcdeg~ and
$d\theta$=10\arcdeg~(the conical curtain). We derived $V_0$ = 190 m/s
(the downrange plume) and $V_0$ = 80 m/s (the conical curtain),
suggesting that the maximum speed was 340 m/s (the downrange plume)  and
140 m/s (the conical curtain) for 0.1-\micron~ particles. As $k$
decreases, the near-nuclear dust cloud 
becomes brighter. The images we captured were consistent with the
simulation images for a value of $k \sim -1/4$. By combining the conical
ejecta curtain and the downrange plume models, we obtained the best-fit
image (Figure ~\ref{fig4}).  Note that we succeeded in the reproduction
of the triple tails by single impact model based on the idea that is
commonly observed in oblique impact experiments. 

As we mentioned above we fixed two variables: $q$ and $a_{min}$.
If there are a large amount of small particles in the dust cloud, it
should look bluer by Rayleigh scattering. As we noticed in
\citet{Ishiguro2011}, no significant difference appears in the
morphology observed in three different optical channels. The
observational evidence implies that the diffuse cloud consisted of dust
particles large enough to scatter optical light (i.e., 2$\pi a /
\lambda$ $>$ 1, where $\lambda$ denotes the optical wavelength). The
size distribution exponent q=-3.5 was applied 
because it is the typical to impact fragment \citep{Dohnanyi1969}. We
performed the fitting with $q$=-4 and $q$=-3, but could not obtain the
plausible result. Therefore, the initial assumptions of $q$ =-3.5 and
$a_{min}$=0.1\micron~ seems to be reasonable. 

In the laboratory experiments, the central axis of the conical curtain
is usually perpendicular to the local surface.  The downrange plume
appears along the trajectory axis of the impactor. Therefore, we can
derive the impactor's trajectory if the impact angle with respect to the
local surface is known. The most probable impact angle on arbitrary
planetary body  is 45\arcdeg \citep{Gault1978}. If
an impact angle of 45\arcdeg~ is assumed, it is likely that a small
asteroid collided with (596) Scheila from the direction of
($\alpha_{im}$, $\delta_{im}$) = (60\arcdeg~, -40\arcdeg~). 
This result suggests that the impactor collided with (596) Scheila from
behind. 
The angle between the central axis of the conical curtain and that of
the downrange plume in the best-fit model is 85\arcdeg, which is potential
value based on the impact experiments (Figure 2(b) in this paper and
Figure 19 of \citet{Schultz2007}).
The derived half-opening angle of 50\arcdeg~  is consistent with the results
obtained for the rocky ejecta with velocity from hundreds to thousands
m/s through the laboratory measurement \citep{Gault1963}.
The power index of size dependence of the ejecta velocity ($k \sim -1/4$)
is smaller than that of comets (i.e. $k\sim1/2$, typical of
hydrodynamical gas drag) but within the range of the laboratory impact
experiments \citep{Giblin1998}.
Our model predicts that up to
140\micron~ particles could escape from (596) Scheila. In fact,
100\micron~ particles were found in the observed images after 2011
February \citep{Ishiguro2011}. 
The velocity of dust particles depends on the tensile strength of the
surface materials when the impact process is dominated by the material
strength rather than gravity. The maximum ejecta velocity for solid and
porous targets measured in the laboratory exceeds 10$\times$
$(Y_t/\rho_t)^{0.5}$, where $Y_t$ is the tensile strength and $\rho_t$ is
the target density (in Figure 18 of \citet{Housen2011} and references
therein), i.e., $v_{max} > 10 \times (Y_t/\rho_t)^{0.5}$. Since the
maximum ejecta velocity is 140 m/s for the conical ejecta, 
$Y_t < \rho_t (v_{max}/10)^2$ $\sim$ 0.3 MPa. 
Our result on the ejecta velocity suggest that the surface on (596)
Scheila was covered by materials with low tensile strength.

\section{Summary}

So far we have outlined a plausible explanation for the peculiar dust
cloud of (596) Scheila. We constructed a model of the morphology based
on experiments of high velocity impacts. The values of the parameters
were free and obtained from fitting the observed image on 2011 December
12. We found that the morphologies on 2011 December 17 and 19
\citep{Ishiguro2011} were also reproduced with the same model
parameters. In summary, we find that:

\begin{enumerate}
\item The morphology of (596) Scheila can be reproduced by an
      impact-driven ejecta plume consisting of an impact cone and
      downrange plume.
\item The maximum ejection velocity of the dust particles exceeded 100
      m/s.
\item Assuming that an impact angle of 45\arcdeg, the impactor collided
      with (596) Scheila from the direction of (60\arcdeg, -40\arcdeg)
      in J2000 coordinates.
\item The surface of (596) Scheila consists of materials with low
      tensile strength ($\sim$0.3 MPa).
\end{enumerate}

With the previous studies
\citep{Jewitt2011,Hsieh2011,Bodewits2011,Ishiguro2011}, we definitively
conclude that a decameter-sized asteroid collided with (596) Scheila
from behind on 2010 December 3. 


\acknowledgments
Research at Seoul National University was supported by the
National Research Foundation of Korea and the Seoul National University
Foundation Research Expense. 
This study was supported by ISAS/JAXA as a collaborative program with
the Space Plasma Experiment.


\clearpage

\clearpage

\clearpage

\clearpage

\clearpage

\begin{table}
\begin{center}
\caption{Input and best-fit parameters for the dust ejection model\label{tbl-1}}
\begin{tabular}{llc}
\tableline\tableline
Parameter & Input values & Best-fit values \\
\tableline
Conical curtain & & \\
$\theta$ [\arcdeg]        & 10 -- 90 with 5 interval & 50 \\
$\delta \theta$ [\arcdeg]  & 1, 3, 5, 10, 15, 20, 30, 40 & 10 \\
$\alpha_{cone}$ [\arcdeg] & 0 -- 360 with 15 interval & 90 \\
$\delta_{cone}$ [\arcdeg] & -90 -- +90 with 10 interval & -15 \\
$V_0$ [m/s] & 60 -- 300 with 10 m/s interval & 80 \\
$k$ & -1/6, -1/5, -1/4, -1/3, -1/2 & -1/4 \\
$q$ & -3.5 (fixed) & -- \\
$a_{min}$ [m] & 1.0$\times$10$^{-7}$ (fixed) & -- \\
$a_{max}$ [m] &	defined as $v_{ej}(a_{max})$=55 m/s & 4$\times$10$^{-6}$ \\
\\
Downrange plume & & \\
$\theta$ [\arcdeg] & 10 -- 90 with 5 interval & 15 \\
$\alpha1_{dwn}$ [\arcdeg] & 0 -- 350 with 15 interval & 150 \\
$\delta1_{dwn}$ [\arcdeg] & -90 -- +90 with 10 interval & +40 \\
$\alpha2_{dwn}$ [\arcdeg] & 10 -- 360 with 15 interval, $\alpha2_{dwn}$
     $>$ $\alpha1_{dwn}$ & 180 \\
$\delta2_{dwn}$ [\arcdeg] & -90 -- +90 with 10 interval & +50 \\
$V_0$ [m/s] & 60 --300 with 10 m/s interval & 190 \\
$k$ & -1/6, -1/5, -1/4, -1/3, -1/2 & -1/4 \\
$q$ & -3.5 (fixed) & -- \\
$a_{min}$ [m] &	1.0$\times$10$^{-7}$ (fixed) & -- \\
$a_{max}$ [m] &	defined as $v_{ej}(a_{max})$=55 m/s & 1.4$\times$10$^{-4}$ \\
\tableline
\end{tabular}
\end{center}
\end{table}

\end{document}